\documentclass[prl, twocolumn, floatfix]{revtex4}
\usepackage{graphicx}
\usepackage{amsmath, amsfonts, amssymb, bm}
\usepackage{braket}
\usepackage{color}
\usepackage[utf8]{inputenc}
\usepackage{mathrsfs}
\usepackage{amsbsy}

\begin{document} 

\title{ Ionization of atoms by dense and compact beams of extreme relativistic electrons } 

\author{ S. Kim, C. M\"uller and A. B. Voitkiv } 
\affiliation{ Institute for Theoretical Physics I, Heinrich-Heine-University D\"usseldorf, 
Universit\"atsstrasse 1, D-40225 D\"usseldorf, Germany } 

\date{\today} 

\begin{abstract}

Ionization is one of the basic physical processes, occurring when 
charged particles penetrate atomic matter. 
When atoms are bombarded by very dense and compact beams 
of extreme relativistic electrons, two qualitatively new -- and very efficient -- ionization mechanisms 
arise: the tunnel or over-barrier ionization and the coherent impact ionization,  
which are driven by the low- and high-frequency parts, respectively, of the beam field. 
In these mechanisms significant fractions of the beam electrons act coherently,  
strongly enhancing the ionization process.    
They are also very sensitive to the spatiotemporal structure of the beam 
that can be used for analysing the beam properties.  
 
\end{abstract} 


\maketitle 

Electron beams are of great importance for many branches of science.  
An especially crucial role is played by such beams in physics, where they find a  
large variety of applications (e.g. for probing subatomic particles, studying fundamental forces  and testing quantum field theories, for probing atomic dynamics on an attosecond time scale, for generating intense, coherent X-rays, in diagnostics of plasma).  

Beams of extreme relativistic electrons, which move with a velocity $v$ practically equal 
to the speed of light $c$, are crucial in cutting-edge physical research \cite{hep}. 
Until recently the densities of high-energy electron beams were relatively low 
and the typical time interval $ \Delta t $ between two consecuitive collisions of the beam electrons 
with the target atom (or nucleus) greatly exceeded 
the typical target transition time $ \tau $.  
As a result,  in collisions with atomic and nuclear targets the beam acted  
as a set of individual electrons whose contributions to target cross sections 
and transition rates add incoherently. 

However, highly relativistic electron beams of unprecedented density 
and up to $\approx 10$ GeV electron energy are nowadays generated by laser or plasma wakefield accelerators \cite{beam1}-\cite{PWFA2}.  
For example, beams of few fs-scale duration with  
currents of $\gtrsim 50$ kA can be produced by laser wakefield accelerators \cite{Coupe},   
and electron beam spikes of $\sim 100$ kA were recently generated   
at the FACET-II facility \cite{Emma}.   
Plasma photocathodes \cite{beam3}-\cite{A1} 
may yield ultrabright electron pulses with sub-fs duration \cite{PC}  
and nanometre-scale normalized emittances, which could be focused to extreme beam densities and corresponding fields of $\sim 10^{10}$ V/cm \cite{NIMA}-\cite{PL3}.  

For such beams the condition $\Delta t \gg \tau$ may no longer be fulfilled. Accordingly, more than one (or even many) electrons can simultaneously interact with an atomic target, representing a qualitatively new physical situation. 

\begin{figure}[t] 
\vspace{-0.05cm}
\begin{center}
\includegraphics[width=0.28\textwidth]{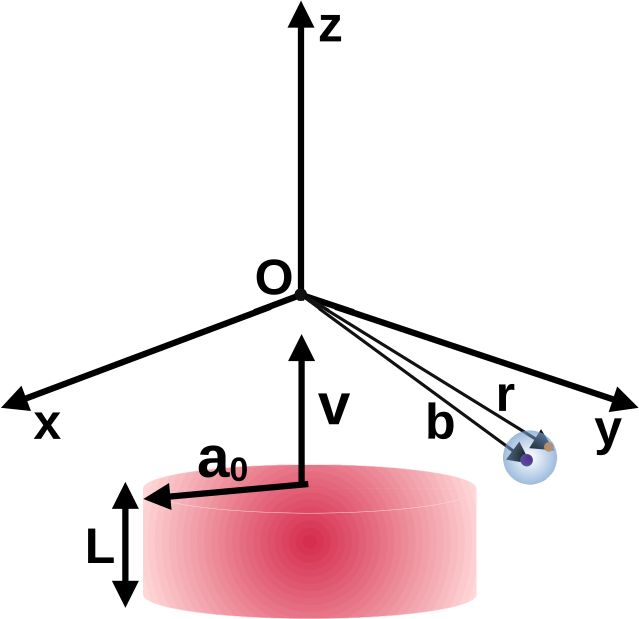} 
\end{center}
\caption{ Sketch of the electron beam - atom collision.   
The vectors $\bm b$ and $\bm r$ denote the coordinates of the atomic nucleus and the atomic electron, respectively. $L$, $a_0$ and $ \bm v = (0,0,v) $ are the beam length, radius and velocity, respectively. The $z$-axis is also the symmetry axis of the beam. } 
\label{figure1}
\end{figure}

In this communication we explore two mechanisms of "collective" 
ionization of atoms by dense beams of extreme relativistic electrons, 
in which significant fractions of these electrons 
coherently interact with the atom  
that may tremendously increase the ionization cross sections. 
In one of them the atom is ionized  
by the low-frequency part of the beam field 
via {\it the tunnel (or over-barrier) ionization} mechanism.
In the other,   
which we shall call {\it coherent impact ionization}, 
the atom is ionized by the interaction with 
the high-frequency part of the beam field. 

It will be seen that both these mechanisms depend very sensitively (but quite differently) 
on the beam parameters.  
This can be exploited for diagnostics of the beams,   
that is important for their envisaged applications, 
such as the generation of coherent x-ray sources \cite{A1}, \cite{A2}-\cite{A3}.

Let the target atoms be initially in the ground state and rest 
in the laboratory frame. In this frame the beam electrons are incident on the target atoms with a velocity $v \approx c$ along the $z$-axis, ${\bf v}=(0,0,v)$, see fig. \ref{figure1}. 
Our consideration will be based on the  
semi-classical approximation, 
in which only the atomic electron(s)  
are treated as quantum particles 
whereas the atomic nucleus  
and the beam electrons are described classically. 
Moreover, it is assumed that the atomic nucleus  
is always at rest and the beam electrons move along straight-line trajectories. 
The former is justified by a 
large nuclear mass (compared to that of the electron) whereas the latter works excellently because the momentum-energy transfers in the collisions are negligibly small compared to the momentum and energy of the incident electrons \cite{q-sc}. 

Let the $ j $-th beam electron  
move along the trajectory $ \bm R_j = {\bm  R }_{\perp, j } + {\bm v} \, (t - t_j) $, where  
$ {\bm  R }_{ \perp, j } $ is its transverse ($ {\bm  R }_{ \perp, j } \perp {\bm v}$) coordinate 
and $t_j$ is the time of its closest approach to the origin. Let the atomic nucleus 
rest at a point with the coordinates $(\bm b, 0)$, where $\bm b = (b_x, b_y) $ is its transverse position vector, and ${\bm r} = (x,y,z)$ be the position of the atomic electron with respect to the origin (see fig. \ref{figure1}).  

The electromagnetic field produced by the $j$-th incident electron 
at the position of the atomic electron 
can be described by the Lienard-Wiechert potentials,   
\begin{eqnarray} 
\Phi_j(\bm r, t) & = & \frac{ \gamma \, e_0 }{\sqrt{ (\bm r_\perp - \bm R_{\perp, j} )^2 + 
\gamma^2 (z - v (t - t_j))^2 }} 
\label{e1}  
\end{eqnarray} 
and $\bm A_j = \bm v/c \, \Phi_j $ (see e.g. \cite{E-M}),  
where $e_0$ is the electron charge, $\gamma = 1/\sqrt{1 - v^2/c^2}$,  
and $\bm r_\perp = (x, y)$.  

Ionization via tunnelling and via absorption of 
high-frequency components from the external field are two facets of 
the same basic process. Nevertheless, they differ qualitatively
and we shall consider them separately, focusing first on   
the (coherent) impact ionization. 

{\bf Impact ionization.} 
In collisions between atoms and high-energy projectiles   
the overwhelming majority of electrons emitted from the atoms have kinetic energies 
not significantly exceeding their initial binding energy 
(see e.g. \cite{ener-dist-e}-\cite{ener-dist-t}). This means that 
in collisions with light atoms, where bound electrons move with velocities $v_0 \ll c$,  
the emitted electrons have kinetic energies $ \varepsilon \ll m_e c^2 $ 
($m_e$ is the electron mass). 
Thus, in such collisions the atomic electron(s) can be described nonrelativistically. 

In collisions with a high-energy electron the effective strength, $\sim e_0^2/\hbar v$, 
of the perturbation acting on the  
atom is very weak. 
Therefore, we shall use the first order of perturbation theory,  
obtaining for the atomic transition amplitude 
\begin{eqnarray} 
S^j_{fi}(\bm q_\perp) \! \! & = & \! \! - \frac{ 2 \, i }{ \hbar } \frac{ e_0^2 }{  v } \, \,   
\, \, e^{ i q_\parallel v t_j }  \, e^{ - i \bm q_\perp \cdot \bm R_{\perp, j} } \, \, 
\frac{ 1 }{ \bm q'^2 }  
\nonumber \\ 
&& \! \! \Big \langle \! \varphi_f \Big \vert e^{i \bm q \cdot \bm \xi} 
- \frac{ v }{ 2 m_e c^2 } \Big( e^{i \bm q \cdot \bm \xi} \hat{p}_z \! + \! \hat{p}_z e^{i \bm q \cdot \bm \xi} \Big) \Big \vert \varphi_i \! \Big \rangle.   
\label{e5}
\end{eqnarray}
Here, $\bm \xi$ is the coordinate of the atomic electron with respect 
to the atomic nucleus, $\varphi_i$ ($ \varepsilon_i$) and $\varphi_f$ ($ \varepsilon_f$) 
are the initial and final states (energies), respectively, 
of the atomic electron, $\omega_{fi} = (\varepsilon_f - \varepsilon_i)/\hbar $ is the atomic transition frequency, $\hat{p}_z = \frac{ \hbar }{ i } \frac{\partial }{ \partial \xi_z } $. 
Further, $ \hbar \bm q = (\hbar \bm q_\perp, \hbar q_\parallel) $ ($\hbar \bm q' = (\hbar \bm q_\perp, \hbar q_\parallel/\gamma) $), where $q_\parallel = \omega_{fi}/v $, is the momentum transferred to the atom in the collision as viewed in the laboratory frame (in the rest frame of the incident electron).

The total transition amplitude is obtained by summing the contributions from all beam electrons,  
\begin{eqnarray} 
S_{fi}(\bm q_\perp) \!\!\! & = & \!\!\! - \frac{ 2 i }{ \hbar } \frac{ e_0^2 }{ v } 
\frac{ \Big \langle \! \varphi_f \Big \vert e^{i \bm q \cdot \bm \xi} \!  
- \! \frac{ v }{ 2 m_e c^2 } \Big( e^{i \bm q \cdot \bm \xi} \hat{p}_z \! + \! \hat{p}_z e^{i \bm q \cdot \bm \xi} \Big) \Big \vert \varphi_i \! \Big \rangle }{ \bm q'^2 } 
\nonumber \\    
&& \times \sum_{j = 1}^{N_t} e^{ i (q_\parallel v t_j - \bm q_\perp \cdot \bm R_{\perp j}) },   
\label{e6}
\end{eqnarray}
where $N_t$ is the total number of these electrons. 

From the structure of the amplitude (\ref{e6}) it is  
obvious that the point of, whether 
more than one electron of the beam may interact coherently with the target,   
depends on the phase factors in the last line of (\ref{e6}). 

Let us introduce the longitudinal and transverse coherence lengths, given by 
$\lambda_\parallel = v / \omega_{fi}$ and $\lambda_\perp = 1/ q_\perp$, respectively 
\cite{f-coherence}. 
$ \lambda_\parallel $ represents the distance, which is traversed by the beam electrons during the atomic transition time $1/\omega_{fi}$.  
Since at $\gamma \gg 1$  
the main contribution to the atomic ionization 
arises in collisions with $q_\perp \ll q_\parallel $  
(see e.g. \cite{we-2001}), 
one has $ \lambda_\perp \gg \lambda_\parallel $. 

Let, further, $\ell_\parallel$ and $\ell_\perp$ be the mean distances between the electrons of the beam in the longitudinal and transverse directions, respectively.   
  
If at least one of the conditions,   
$ \lambda_\parallel \ll \ell_\parallel $ and/or  
$ \lambda_\perp \ll \ell_\perp $ is met, the double sum over the beam electrons in 
$ \vert S_{fi}(\bm q_\perp) \vert^2 $ (arising from the last line of Eq. (\ref{e6}))  
reduces to the total number $ N_t $ of the electrons in the beam 
due to their random positions in space. In such a case the beam interacts with the atom as 
an incoherent set of individual electrons. This situation represents the "normal" 
regime in high-energy atomic collisions 
(and in high-energy physics in general) 
and is well studied. In particular, the total cross section $\sigma_t$ 
for single ionization by the impact of $N_t$ individual electrons behaves 
at high energies as       
\begin{eqnarray} 
\sigma_t = N_t \, \, \frac{ \alpha_1  }{ v^2 } \, \, \Big (\ln( \alpha_2 \, v \, \gamma ) 
- 0.5 \, v^2/c^2\Big),  
\label{e7} 
\end{eqnarray}
where $\alpha_1$ and $\alpha_2$ depend solely on the target atom   
(e.g. for single ionization of He(1s$^2$) $\alpha_1 = 12.289$ and $\alpha_2 = 2.08 $ \cite{He}, 
if atomic units,  
$\hbar = |e_0| = m_e = 1$, are used). 
   
A qualitatively different collision regime arises when the conditions  
$ \lambda_\parallel \gtrsim \ell_\parallel $ and  $ \lambda_\perp \gtrsim \ell_\perp $ 
are fulfilled. Then the beam no longer acts 
as a set of individual electrons. Instead, now many electrons from the beam simultaneously 
interact with the atom, which "sees" the beam essentially as a continuous charge distribution. 
In such a case the last line in Eq. (\ref{e6}) is transformed according to  
\begin{eqnarray} 
\sum_{j = 1}^{N_t} e^{ i (q_\parallel v t_j - \bm q_\perp \cdot \bm R_{\perp j}) } 
\to \int d^3 {\bm R } \, \, \rho( \bm R, t) \, \, e^{ i (q_\parallel Z  - \bm q_\perp \cdot \bm R_{\perp}) } 
\label{e8}
\end{eqnarray} 
where $ \bm R = (\bm R_\perp, Z) $, $ \rho( \bm R, t) $ 
is the beam density and the integration runs over the 
volume occupied by the beam. 

The total cross section for impact ionization $\sigma_{\rm ion}$ reads 
\begin{eqnarray} 
\sigma_{\rm ion} = \int \frac{ d^3 {\bm p} \, V }{ (2 \pi \hbar)^3  }  
\int d^2 \bm q_\perp  \, \, \vert S_{fi}(\bm q_\perp) \vert^2,      
\label{e10}
\end{eqnarray} 
where $ {\bm p} $ and $V$ are the momentum and the normalization volume, respectively, 
of the emitted electron. 

\begin{figure}[t] 
\vspace{-0.15cm}
\begin{center}
\includegraphics[width=0.3\textwidth]{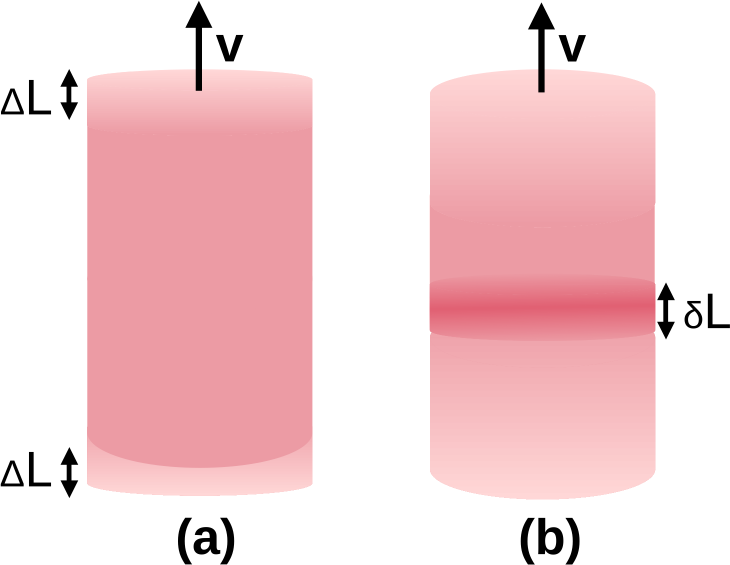}
\end{center}
\vspace{-0.5cm}
\caption{ Sketch of a beam whose intensity: a) "turns on and off"       
within $ \Delta L / v $,  
b) has a spike with duration $\delta L / v $.  } 
\vspace{-0.37cm}
\label{figure2}
\end{figure}

As in ionization by individual projectiles, the coherent impact ionization 
can be viewed as occurring via the absorption of  
'equivalent photons' \cite{weiz-willi, jack}, 
generated in this case by the concerted action of the beam electrons.  
Therefore, for this mechanism to be efficient, there must be enough photons 
with energies greater than the atomic binding energy. This holds if:  
 
i) the beam is short, $ L \lesssim 10 \, v \, \tau  $,   
where $ \tau $ is the effective target 
transition time;   

ii) the beam is long, but along the propagation direction its density  
significantly varies over distances $ \lesssim 10 \, v \, \tau $ 
(e.g., in its front and/or rear part, or in spikes, see fig. \ref{figure2} 
for an illustration).  

{\bf Tunnel/over-barrier ionization.} When the field, coherently created by the 
beam electrons, varies slowly on the atomic time scale but is sufficiently strong, 
the atom is ionized via tunnelling \cite{LL-2}, \cite{ADK}, \cite{bm}. For even stronger fields, 
which are comparable to the internal atomic field or even exceed it, 
the over-barrier ionization occurs. 
The tunnel and over-barrier ionization are highly nonperturbative processes,  
qualitatively different from 
the (perturbative) impact ionization, 
which requires 
sufficiently high frequency components 
in the projectile's field but in general does not necessitate  
strong fields. 

{\bf Ionization versus atomic position.} An additional insight into 
the ionization process is obtained by exploring the atomic ionization 
probability $ P_{\rm ion}(b) $ as a function of the atomic position $b$ 
with respect to the beam axis (see fig. \ref{figure1}). 
(We note that the ionization cross sections 
are given by $2 \pi \int_0^\infty  db \, b \, P_{\rm ion}(b) $.) 
 
\begin{figure}[t] 
\vspace{-0.35cm}
\begin{center}
\includegraphics[width=0.35\textwidth]{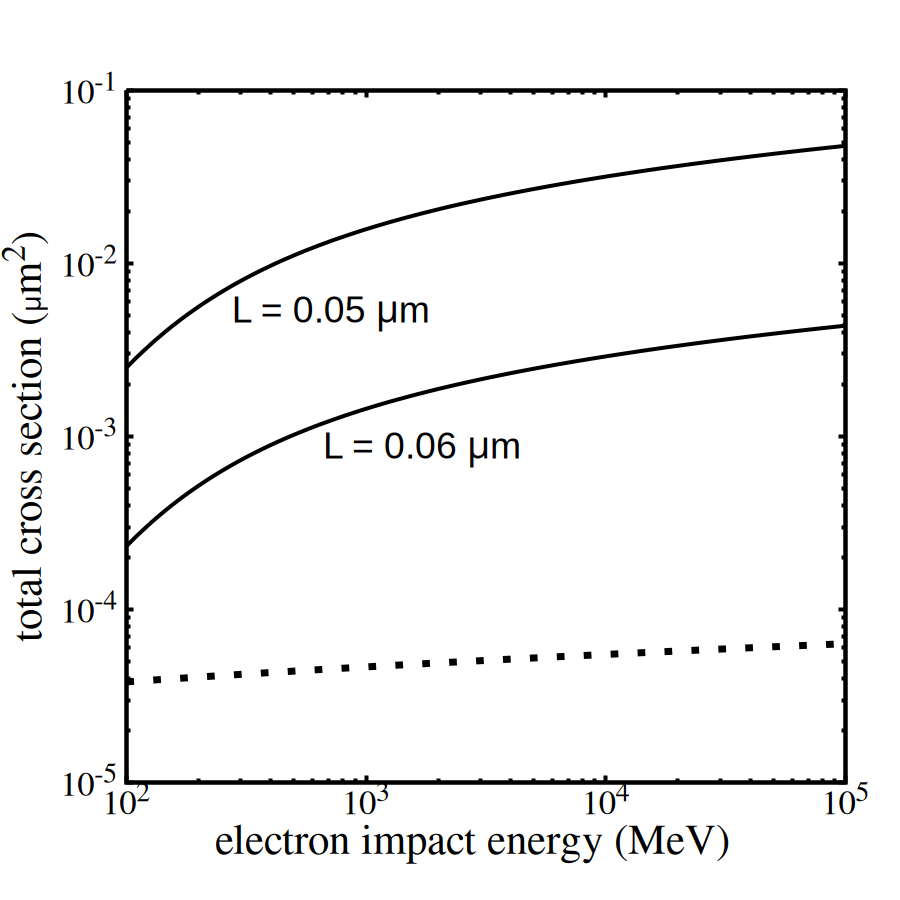}
\end{center} 
\vspace{-0.95cm}
\caption{ Single ionization of He(1s$^2$) atoms  
by electron beams, as a function 
of the electron  
energy. $N_t = 2 \times 10^6$, $a_0 = 1 $ $\mu m$, 
$L = 0.05 $ $\mu m$ and $0.06 $ $\mu m$, as indicated.  
Solid curves: the coherent impact ionization. 
Dot curve: the cross section (\ref{e7}). } 
\vspace{-0.5cm}
\label{figure3}
\end{figure}

\begin{figure}[t] 
\vspace{-0.1cm}
\begin{center}
\includegraphics[width=0.35\textwidth]{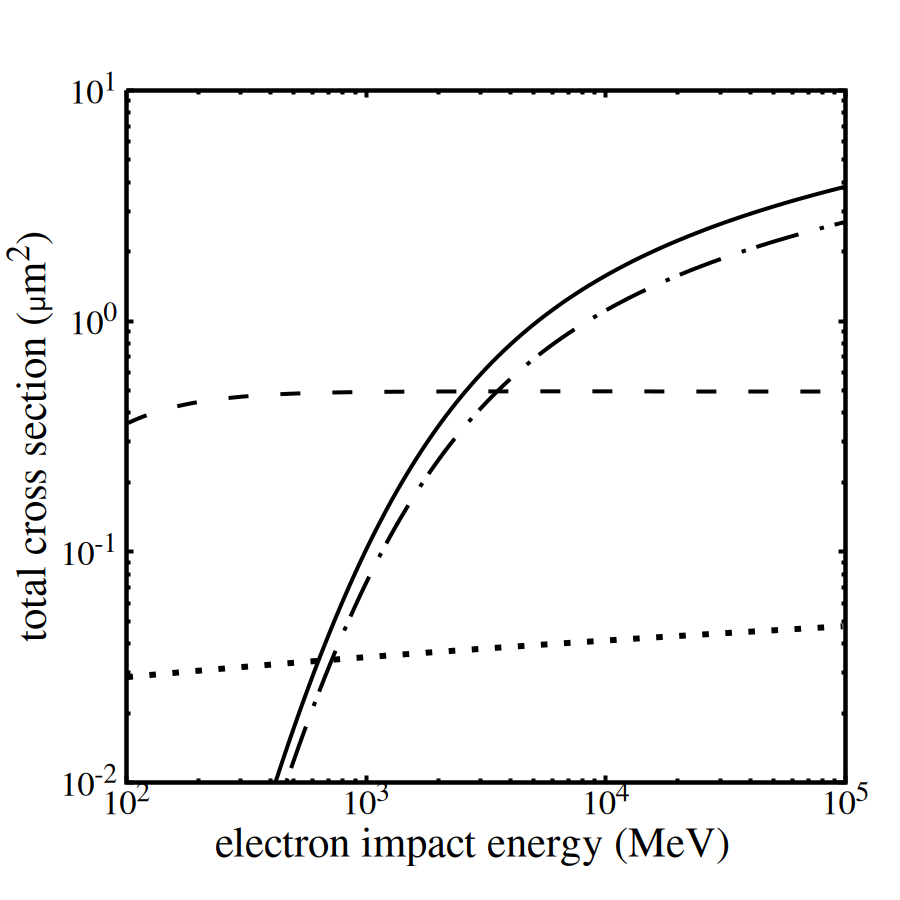}
\end{center}
\vspace{-0.95cm}
\caption{ Single ionization of He(1s$^2$)   
by beams with $N_t = 1.5 \times 10^9$, $L = 3 $ $\mu m$ ($L/v = 10 $ fs), $a_0 = 20$ $\mu m$. 
Solid and dash-dot curves: the coherent impact ionization. 
Dash curve: tunnel ionization. 
Dot curve: the cross section (\ref{e7}).  
For more explanations see text.  } 
\vspace{-0.37cm}
\label{figure4}
\end{figure}

{\bf Numerical results and discussion. }   
In figs. \ref{figure3} and \ref{figure4} we show the results for the total cross sections 
of single ionization of He(1s$^2$), when a helium gas target is penetrated 
by beams with electron energy ranging 
between $ 100 $ MeV and $ 100 $ GeV ($ 2 \times 10^2 \lesssim \gamma  \lesssim 2 \times 10^5 $).  The cross sections are given {\bf per beam}. 

The results for the coherent impact ionization,  
shown in fig. \ref{figure3}, were obtained by taking the beam density as  
\begin{eqnarray} 
\rho( \bm R, t) = \frac{ 2 \,  N_t }{ \pi^{3/2} \, \, a_0^2 \, L   } \,  \, 
e^{- \eta^2/(L/2)^2 }  \, \, 
e^{ - R_\perp^2/a_0^2},     
\label{e9}
\end{eqnarray} 
where $N_t = 2 \times 10^6 $, $\eta = Z - vt $ and $a_0 = 1 $ $ \mu m$. 
Two short beam lengths,  
$ 0.05 $ $\mu m$ and $ 0.06 $ $\mu m$, are considered 
(the associated beam currents are $\lesssim 2$ kA).    
The beam pulse time, $ L/v = 0.167 $ fs and $ 0.2$ fs, 
can be compared to the effective target transition time $\tau \approx 0.02 $ fs 
in the helium impact ionization. We also note that for these short beams the tunnel ionization is negligible. 

In fig. \ref{figure4} we present the results for single ionization of He(1s$^2$)   
by two long beams, each with $N_t = 1.5 \times 10^9$, $L = 3 $ $\mu m$  
and $a_0 = 20$ $\mu m$. (The associated beam currents are $ \simeq 24$ kA.)  
The density of the first beam reads 
\vspace{-0.05cm}
\begin{equation} 
\rho \! = \! \frac{N_t  e^{- \frac{ R_\perp^2 }{ a_0^2 } } }{ \pi a_0^2 L } \! \left \{ \!\!\!\!\! 
\begin{array}{rl}  
&\! 0 ; \eta \leq - \frac{\Delta L}{ 2 } \,\, \, {\rm or} \, \, \,  \eta \geq L +  \frac{\Delta L}{ 2 } \\ 
& \! \frac{ 1 }{ 2 } \Big( \! 1 \! + \! \sin \Big(\!\frac{ \pi \eta }{ \Delta L } \!\Big) \!\Big) ; - 
\frac{\Delta L}{2} \leq \eta \leq \frac{\Delta L}{2}  \\  
& \! 1;  \, \, \frac{\Delta L}{2} \leq \eta \leq L - \frac{\Delta L}{2} \\ 
& \! \frac{ 1 }{ 2 } \Big(\! 1 \! - \! \sin \Big(\!\frac{ \pi (\eta - L) }{ \Delta L} \!\Big) \!\Big) ; 
L \! - \! \frac{\Delta L}{2} \! \leq \! \eta \leq L \! + \! \frac{\Delta L}{2}  .    
\end{array}  
\right.
\label{beam1} 
\end{equation} 
The form of the density profile of the second beam is basically 
as in Eq. (\ref{e9}) but, in addition,  
this beam has a spike with the density 
$ \delta \rho \sim e^{- 4 \eta^2/(\delta L)^2} \, e^{- R_\perp^2/a_0^2} $,  
containing $2 \times 10^{7}$ electrons.

The cross sections for the coherent impact ionization by the first and second beam 
with $\Delta L = \delta L = 0.05$ $\mu m$  
are shown in fig. \ref{figure4} by 
the solid and dash-dot curves, respectively. The figure also displays  
the cross section for the tunnel/over-barrier ionization 
which is the same for both beams, and the cross section (\ref{e7}) for 
ionization by individual electrons. 

The results displayed in figs. \ref{figure3} and \ref{figure4} demonstrate 
that, when the condition i) or ii) is fulfilled, the coherent impact ionization 
is an efficient mechanism which can strongly outperform the tunnel  
ionization and the ionization by individual electrons.  

The latter process depends weakly ($ \sim \ln \gamma$) on the impact energy, 
as is inherent to the processes of ionization and bound-free $ e^+  e^-$ pair production 
(see e.g. \cite{abv_review}) in high-energy collisions with individual  
projectiles.
 
The tunnel cross section 
at sufficiently high energy is practically constant. This  
can be understood by noting that at very high impact energies 
the low-frequency part of the beam field and the beam velocity 
are practically independent of the impact energy. Therefore, 
neither the tunnelling rate nor the time interval, when the tunnelling can occur, 
noticeably vary with the impact energy.   

\begin{figure}[t] 
\vspace{-0.35cm}
\begin{center}
\includegraphics[width=0.35\textwidth]{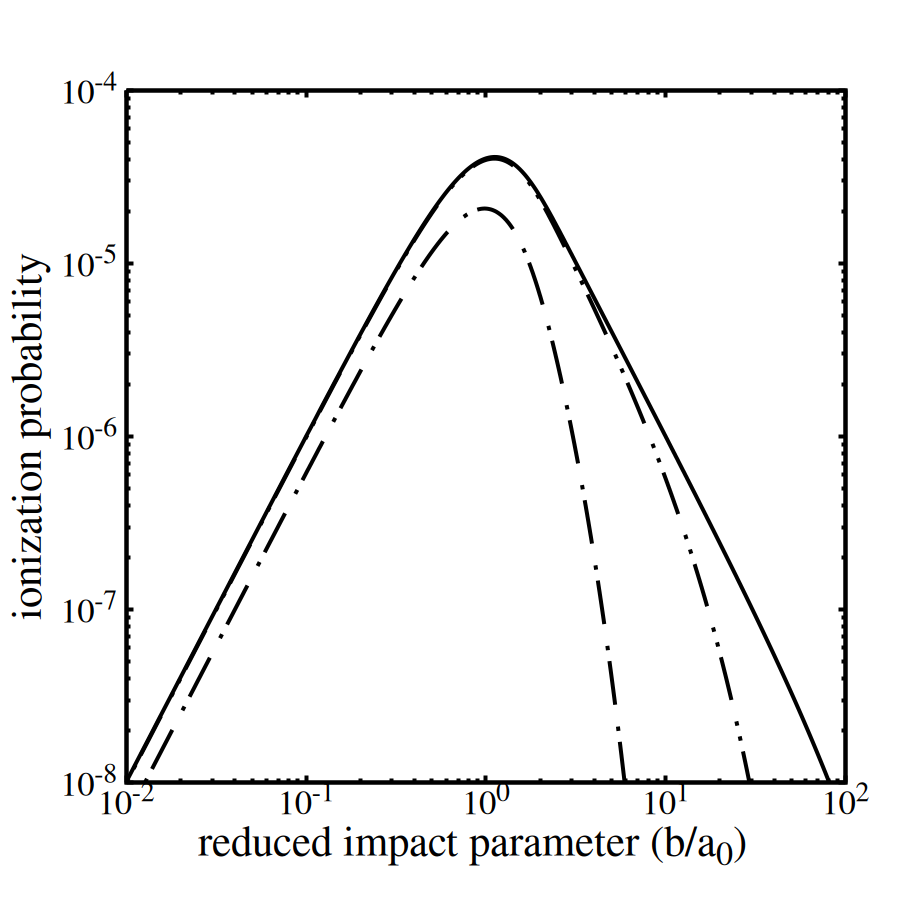}
\end{center} 
\vspace{-0.95cm}
\caption{ The probability $ P_{\rm ion}(b) $ for single ionization of He(1s$^2$) 
as a  function of the position $ b $ 
of the atom with respect to the beam axis 
(see fig. \ref{figure1}).  
$N_t = 2 \times 10^6$, $L = 0.06 $ $\mu$m, $a_0 = 1 $ $\mu$m. 
Dash-dot, dash-dot-dot and solid curves: 
coherent impact ionization by 100 MeV, 1 GeV and 10 GeV beams, respectively.  
} 
\vspace{-0.5cm}
\label{figure5}
\end{figure}

In contrast,  
the cross section for the coherent impact ionization 
depends on the energy rather strongly.   
In order to get insight into the origin of this dependence 
let us consider the corresponding ionization probability. 
In fig. \ref{figure5} this probability is shown for single ionization of He(1s$^2$) 
by a gaussian beam (see Eq. (\ref{e9})) with $N_t = 2 \times 10^6$, $L = 0.06 $ $\mu$m, $a_0 = 1 $ $\mu$m at 
electron impact energies of $100$ MeV, $1$ GeV and $10$ GeV. 
It is seen that the probability increases with the impact energy:   
it especially strongly extends towards larger $b$, indicating 
an increase in the interaction range between the beam electrons 
and the target atoms.   

In collisions with individual charged projectiles the effective range of 
the projectile-atom interaction is determined by the adiabatic collision radius 
$ R_0 \simeq v \, \gamma \, \tau $. 

In collisions with the beam  
the growth of $R_0$ leads i) to an increase of the interaction range for a particular 
electron-atom pair and, in addition, ii) to an increase of the number of the beam electrons coherently 
involved in the generation of equivalent photons. 
The point ii) is obviously limited since 
the "stock" of the electrons available will eventually be exhausted. 
Indeed, our analysis shows that at asymptotically high impact energies 
($ \gamma \gg a_0/c\tau$)  
the cross section behaves as $\sim \ln \gamma $, i.e. its energy dependence becomes 
similar to that in collisions with individual projectiles.   

Our analysis shows that the coherent impact ionization 
is very sensitive to the beam parameters (see for an illustration fig. \ref{figure3}) 
and that the same is true also for the tunnel ionization. 
However, while the coherent impact ionization is especially 
sensitive to the longitudinal profile of the beam (which determines the energy distribution of the equivalent photons), the tunnelling very strongly depends  
on its transverse profile and the quantity $N_t / L$ (both influence the beam field strength $F$ 
to which the tunnelling ionization rate  
$\sim \frac{F_a}{F} \exp(-2 F_a / 3 F) $, where $F_a \sim 10^9$ V/cm is the characteristic atomic field, is extremely sensitive). 

{\bf Target size and screening effects. }  
The growth of the impact cross section(s) with the collision energy will eventually 
be limited either by the transverse size of the target 
(we assume that it is much larger than the beam radius $a_0$)  
and/or by the target density effect, 
in which atoms closer to the projectile trajectory shield the projectile field for atoms farther away. Both these effects were omitted in our consideration which is justified when the transverse size of the gas target greatly exceeds the adiabatic collision radius $R_0$ and the target density is sufficiently low (see e.g. \cite{jack}, \cite{LL-8}). 
At the impact energies, considered in figs \ref{figure3} and \ref{figure4}, $ R_0 \lesssim 1 $ mm. Thus, the target size effect will be unimportant if $d_{\rm tr} \gtrsim 5$ mm, and, as our estimates show, even at $100 $ GeV the cross sections will still be unaffected by the density effect, if the helium density $\lesssim 10^{13}$ cm$^{-3}$.    

{\bf Ionization by secondary particles. } 
The atoms can also be ionized by the electrons,  
emitted due to the interaction with the beam, 
and by the bremsstrahlung radiation. 
However, our estimates show that the former  
can be neglected for helium densities   
$\lesssim 10^{17}$ cm$^{-3}$, whereas the latter is 
negligible for any possible helium density.

{\bf In conclusion}, we have explored two "collective" mechanisms 
of atomic ionization by high-density compact beams of extreme relativistic electrons,   
in which siginificant fractions of these electrons  
coherently interact with the atom that enhances the ionization.    
As a result, the corresponding cross sections   
can strongly exceed  
the ionization cross section 
obtained by assuming that the beam electrons act individually.    

The coherent ionization mechanisms possess interesting 
peculiarities, including a very high sensitivity to 
the spatiotemporal structure of the beam, which makes 
their account necessary 
for analysing the beam properties. 

\section*{Acknowledgement} 

We thank B. Najjari, B. Hidding and T. Heinemann for useful discussions. 
S.K. gratefully acknowledges funding by the Studienstiftung des deutschen Volkes.


\begin{thebibliography}{99}

\bibitem{hep} V. Shiltsev, F. Zimmermann,  
Reviews of Modern Physics 
{\bf 93}, 015006 (2021).   

\bibitem{beam1}  
E. Esarey, C. B. Schroeder, and W. P. Leemans, 
Reviews of Modern Physics 
{\bf 81}, 1229 (2009).  

\bibitem{beam2} T. Tajima, X. Q. Yan, T. Ebisuzaki, 
Reviews of Modern Plasma Physics (2020) 4:7 

\bibitem{PWFA1} E. Gschwendtner and P. Muggli, 
Nature Rev. Phys. {\bf 1}, 246 (2019). 

\bibitem{PWFA2} C. A. Lindstrøm, S. Corde, R. D'Arcy, S. Gessner, M. Gilljohann, M. J. Hogan, and J. Osterhoff, 
arxiv.2504.05558 .  

\bibitem{Coupe} J. P. Couperus, R. Pausch, A. K\"ohler,  O. Zarini, J. M. Kr\"amer, M. Garten, A. Huebl, R. Gebhardt, U. Helbig, S. Bock, K. Zeil, A. Debus, M. Bussmann, U. Schramm, and A. Irman,
Demonstration of a beam loaded nanocoulomb-class laser wakefield accelerator,
Nat. Commun. {\bf 8}, 487 (2017).  

\bibitem{Emma} C. Emma, N. Majernik, K. K. Swanson, R. Ariniello, S. Gessner,     
R. Hessami, M. J. Hogan, A. Knetsch, K. A. Larsen, A. Marinelli,  
B. O’Shea, S. Perez, I. Rajkovic, R. Robles, D. Storey, and G. Yocky,
Phys. Rev. Lett. {\bf 134}, 085001 (2025). 

\bibitem{beam3} A. F. Habib, Th. Heinemann, G. G. Manahan, D. Ullmann, P. Scherkl, 
A. Knetsch, A. Sutherland, A. Beaton, D. Campbell, L. Rutherford, et al., 
Ann. Phys. (Berlin), {\bf 535}, 2200655 (2023).  

\bibitem{PC} A. Deng, O. S. Karger, T. Heinemann, A. Knetsch, P. Scherkl, G. G. Manahan, A. Beaton, D. Ullmann, G. Wittig, A. F. Habib, et al., 
Nature Physics {\bf 15}, 1156 (2019). 

\bibitem{A1} A. F. Habib, G. G. Manahan, P. Scherkl, T. Heinemann, A. Sutherland, R. Altuiri, B. M. Alotaibi, M. Litos, J. Cary, T. Raubenheimer, et al., 
Nat. Commun. {\bf 14}, 1054 (2023). 

\bibitem{NIMA} J.B. Rosenzweig, G. Andonian, P. Bucksbaum, M. Ferrario, S. Full, 
A. Fukusawa, E. Hemsing, B. Hidding, M. Hogan, P. Krejcik, P. Muggli, G. Marcus, A. Marinelli, 
P. Musumeci, B. O’Shea, C. Pellegrini, D. Schiller, G. Travish, 
Nuclear Instruments and Methods in
Physics Research {\bf A 653}, 98 (2011); 

\bibitem{accel-beams} S. Kuschel, D. Hollatz, T. Heinemann, O. Karger,  M. B. Schwab, D. Ullmann, 
A. Knetsch, A. Seidel,  C. R\"odel, M. Yeung, M. Leier, A. Blinne, H. Ding, T. Kurz, 
D. J. Corvan, A. Sävert, S. Karsch, M. C. Kaluza, B. Hidding, and M. Zepf,  
Phys. Rev. Accel. Beams {\bf 19}, 071301 (2016). 

\bibitem{PL1} C. Thaury, E. Guillaume, A. D\"opp, R. Lehe, A. Lifschitz, K. Ta Phuoc, J. Gautier, J.-P. Goddet, A. Tafzi, A. Flacco, et al., 
Nat. Commun. {\bf 6}, 6860 (2015). 

\bibitem{PL2} C. E. Doss, E. Adli, R. Ariniello, J. Cary, S. Corde, B. Hidding, M. J. Hogan, K. Hunt-Stone, C. Joshi, K. A. Marsh, et al., 
Phys. Rev. Accel. Beams {\bf 22}, 111001 (2019). 

\bibitem{PL3} Y.-Y. Chang, J. Couperus Cabadağ, A. Debus, A. Ghaith, M. LaBerge, R. Pausch, 
S. Sch\"obel, P. Ufer, U. Schramm, and A. Irman, 
Phys. Rev. Applied {\bf 20}, L061001 (2023). 

\bibitem{A2} E. G. Gelfer, A. M. Fedotov, O. Klimo, and S. Weber, 
Phys. Rev. Research {\bf 6}, L032013 (2024). 

\bibitem{A3} M. J. Quin, A. Di Piazza and M. Tamburini, 
Plasma Phys. Control. Fusion {\bf 67}, 055008 (2025). 

\bibitem{q-sc} 
We have also performed calculations by describing beam electrons  
by Dirac plane-wave states. After the averaging over spin states of the incident electron and summing over spin states of the scattered electron (taking into accout that the scattering angle is very small), the resulting cross sections fully coincide with those which are obtained 
by using the semi-classical approach. 

\bibitem{E-M} 
J. Eichler, {\it Lectures on Ion-Atom collisions} (Elsevier, New York 2005).  
 
\bibitem{ener-dist-e} 
R. Moshammer, W. Schmitt, J. Ullrich, H. Kollmus, A. Cassimi, 
R. D\"orner, O. Jagutzki, R. Mann, R. E. Olson et al, 
Phys. Rev. Lett. {\bf 79}, 3621 (1997). 

\bibitem{ener-dist-t} 
A. B. Voitkiv, B. Najjari, R. Moshammer 
and J. Ullrich, Phys. Rev. {\bf A 65}, 032707 (2002);    
A. B. Voitkiv and B. Najjari, 
J. Phys. {\bf B 37}, 4831 (2004).

\bibitem{f-coherence} $\lambda_\parallel $ and $\lambda_\perp$ can be viewed as 
the longitudinal and transverse wavelengths, respectively, of the equivalent photon.  

\bibitem{we-2001} A. B. Voitkiv and J. Ullrich, J. Phys. {\bf B 34}, 4513 (2001). 

\bibitem{He} M. Inokuti, Rev. Mod. Phys. {\bf 43}, 297 (1972); 
Rev. Mod. Phys. {\bf 50}, 23 (1978). 

\bibitem{weiz-willi} 
E. Fermi, Z. Phys. {\bf 29}, 315-327
(1924); C. F. von Weizs\"acker, Z. Phys. {\bf 88}, 612 (1934); 
E. J. Williams, Kgl. Danske Videnskab. Selskab
Mat.-fys. Medd. {\bf 13}, No. 4 (1935). 

\bibitem{jack} J.D. Jackson, {\it Classical Electrodynamics}, 3rd ed., (Wiley,
New York, 1999).

\bibitem{LL-2} L.D. Landau and E.M. Lifshitz, {\it Quantum Mechanics} 
(Pergamon, New York, 1965).     

\bibitem{ADK} M. V. Ammosov, N. B. Delone, and V. P. Krainov, Zh.
Eksp. Teor. Fiz. {\bf 91}, 2008 (1986) [Sov. Phys. JETP {\bf 64},
1191 (1987)].

\bibitem{bm}G. G. Manahan, A. Deng, O. Karger, Y. Xi, A. Knetsch, M. Litos, G. Wittig, 
T. Heinemann, J. Smith, Z. M. Sheng, D. A. Jaroszynski, G. Andonian, 
D. L. Bruhwiler, J. B. Rosenzweig, and B. Hidding. 
Phys. Rev. Accel. Beams {\bf 19}, 011303 (2016). 
%
 
\bibitem{abv_review}  
A. B. Voitkiv, {\it Basic Atomic Processes in High-Energy Ion–Atom
Collisions}, Chapter 57 in { Springer Handbook of
Atomic, Molecular, and Optical Physics}, 2nd Edition, 
ed. by Gordon W. F. Drake (2023). 

\bibitem{LL-8} L.D. Landau and E.M. Lifshitz, 
{\it Electrodynamics of Continuous Media} (Pergamon Press, New York, 1984), 
see § 85.   

\end{thebibliography}
\end{document}